\PassOptionsToPackage{table}{xcolor}
\documentclass{article} %
\usepackage{preprint,times}
\usepackage{booktabs}
\usepackage{array}
\usepackage[table]{xcolor}

\usepackage{amsmath,amsfonts,bm}

\def\eqref#1{equation~\ref{#1}}

\def\1{\bm{1}}

\DeclareMathAlphabet{\mathsfit}{\encodingdefault}{\sfdefault}{m}{sl}
\SetMathAlphabet{\mathsfit}{bold}{\encodingdefault}{\sfdefault}{bx}{n}

\PassOptionsToPackage{table}{xcolor}
\usepackage{booktabs}
\usepackage{amsfonts}
\usepackage{graphicx}
\usepackage{hyperref}
\usepackage{textcomp}
\usepackage{listings}
\usepackage{adjustbox}
\usepackage{xspace}    %
\usepackage{multirow}
\usepackage{multicol}
\usepackage{xcolor}
\usepackage{amsmath}
\usepackage{tcolorbox}
\usepackage[linesnumbered,boxed,ruled]{algorithm2e}
\usepackage{listings}
\usepackage{pifont}
\usepackage{enumitem}

\usepackage{wasysym}
\usepackage{algorithmic}

\definecolor{codegreen}{rgb}{0,0.6,0}
\definecolor{codegray}{rgb}{0.5,0.5,0.5}
\definecolor{codepurple}{rgb}{0.58,0,0.82}
\definecolor{backcolour}{rgb}{0.95,0.95,0.92}

\lstdefinestyle{mystyle}{
  backgroundcolor=\color{backcolour},   commentstyle=\color{codegreen},
  keywordstyle=\color{magenta},
  numberstyle=\tiny\color{codegray},
  stringstyle=\color{codepurple},
  basicstyle=\ttfamily\footnotesize,
  breakatwhitespace=false,
  breaklines=true,
  captionpos=b,
  keepspaces=true,
  numbers=left,
  numbersep=5pt,
  showspaces=false,
  showstringspaces=false,
  showtabs=false,
  tabsize=2
}

\usepackage{nicematrix}

\usepackage{amsthm}
\theoremstyle{definition}

\usepackage{tikz}
\usepackage{fvextra}
\DeclareUnicodeCharacter{2192}{\ensuremath{\rightarrow}}

\usepackage{hyperref}
\usepackage{url}

\usepackage{caption}
\usepackage{booktabs}
\usepackage{multirow}
\usepackage{colortbl}
\usepackage[table]{xcolor}
\usepackage{float}

\definecolor{rowgray}{gray}{0.93}

\usepackage{booktabs}
\usepackage{multirow}
\usepackage{colortbl}
\usepackage[table]{xcolor}
\usepackage{float}

\usepackage{soul}
\usepackage{xcolor}

\sethlcolor{yellow}

\usepackage{booktabs}
\usepackage{tabularx}
\usepackage{subcaption}

\title{StateHarness: Maintaining Evolving Project State for Long-Horizon Coding Agents}

\author{
Chuan Zou$^{1*}$ \quad
Yujun Zhou$^{2*}$ \quad
Xiaonan Luo$^{2}$ \quad
Xiangliang Zhang$^{2\dagger}$ \\
$^{1}$University of Pennsylvania \quad
$^{2}$University of Notre Dame 
\\
$^{*}$Co-first authors \quad $^{\dagger}$Corresponding author
}

\begin{document}

\maketitle

\begin{abstract}
Long-horizon coding requires agents to maintain an evolving view of the
project they are changing: what problem is currently being solved, what has
already been learned or modified, which other artifacts may be affected, and
what work remains unresolved. When this information is scattered across a
growing interaction history, local edits can become disconnected from their
later consequences and progress can be difficult to carry forward.
We introduce \textbf{StateHarness}, a runtime framework that maintains
evolving project state without updating model parameters. The working agent
records its current problem and hypothesis, important findings and decisions,
discovered repository relationships, and unfinished work. Runtime support
prompts record updates, searches for callers, tests, and documentation that
may be affected by a change, and maintains a compact working brief for
subsequent decisions. When behavioral signals accumulate, a separate
diagnostic session revisits the recorded history and surfaces overlooked
evidence.
Across three repository-level evaluations, StateHarness improves task
completion and several measures of change consistency. On RefactorBench,
full-chain completion increases from 29.4\% to 35.3\% and cleanup-check
accuracy from 87.0\% to 95.4\%. On five LHTB software-engineering tasks,
mean reward improves from 0.427 to 0.576, exceeding LongHorizon-Harness at
substantially lower recorded execution cost. On ChainSWE, bug-fix rate
increases from 35.1\% to 39.5\% and residual debt decreases from
60 to 43 items, although pass-to-pass regressions increase. These results
show that maintaining project state can improve long-horizon coding while
introducing measurable execution and regression trade-offs.
Code is available at
\url{https://anonymous.4open.science/r/StateHarness}.
\end{abstract}
\section{Introduction}
\label{sec:introduction}

Recent advances in large language models have moved automated programming
from isolated code generation toward coding agents that interact with
development environments, invoke tools, and carry out multi-step software
engineering tasks~\citep{yao2023react,yang2024sweagent,openhands}. These agents
now operate directly in real repositories, while the tasks themselves are
expanding from issue resolution to complex multi-file changes, feature
development, and sustained software evolution~\citep{jimenez2024swebench,
deng2026swebenchpro,zhou2026featurebench,thai2025swevo,lam2026swechain}.

This shift exposes a mismatch between how agents work and how software projects
are organized. An agent experiences a task as a trajectory of file reads,
edits, tests, and observations. A software project, however, is a structured
and evolving system: modules serve different roles, implementations interact
across files and layers, and a local change can affect callers, tests,
documentation, or other dependent artifacts many steps later. Without a
persistent project-level view, the agent may repeatedly reconstruct the same
understanding from raw code, lose track of the wider consequences of a local
edit, and scatter progress, decisions, and unfinished work across a growing
trajectory. Recent evaluations expose related failures, including forgotten
requirements in multi-turn coding~\citep{raghavendra2026sweinteract},
reduced implementation faithfulness when specifications emerge
gradually~\citep{yan2026projectguard}, and structural degradation under
repeated code extension~\citep{orlanski2026slopcodebench}. Simply retaining
more history is not enough: prior information helps only when the relevant
parts can be selected and reused appropriately~\citep{zhu2026swecontextbench}.

\begin{figure}[t]
\centering
\includegraphics[width=\linewidth]{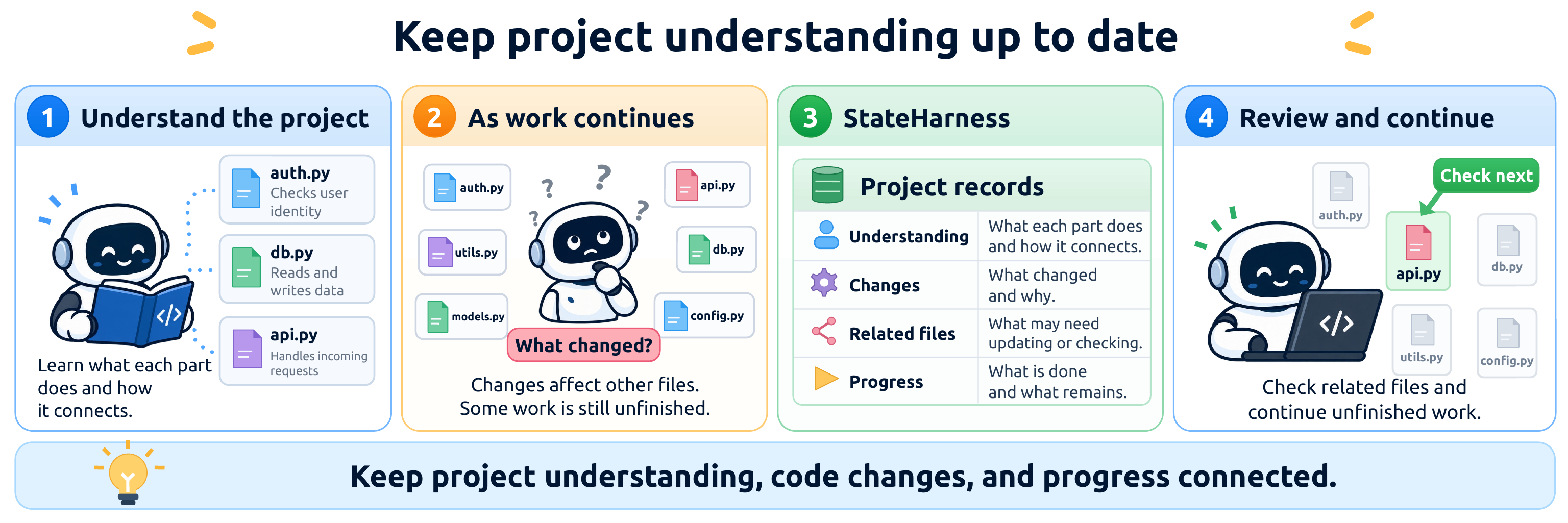}
\caption{Maintaining evolving project state as coding progresses.
StateHarness preserves what the agent has learned, records code changes and
potentially related files, and tracks what has been completed and what still
requires attention. These records support reviewing change consequences and
continuing unfinished work.}
\label{fig:project-state}
\end{figure}

Existing work supports long-horizon coding through context management,
persistent execution state, and repository-aware retrieval
~\citep{liu2026contexttool,gao2026swemem,wang2026ledger,lh_harness,
ouyang2025repograph}. These approaches make different parts of prior execution
easier to retain or recover. We focus on how an evolving, task-centered project
state can keep these pieces connected during ongoing coding: what the agent
currently understands about the project, what consequences recent changes may
have elsewhere, and what work remains unresolved as the repository changes.

We introduce \textbf{StateHarness}, a runtime framework for maintaining an
evolving, task-centered project state throughout execution
(Figure~\ref{fig:project-state}). The agent persistently records its current
problem and hypothesis, important findings and decisions, discovered relations
across the repository, and unfinished work. Runtime support prompts these
records to be updated, searches for callers, tests, and documentation that may
be affected by a change, and maintains a compact working brief for subsequent
decisions. When behavioral signals accumulate, a separate diagnostic session
can revisit the longer recorded history and surface overlooked evidence.
StateHarness operates without updating the underlying model parameters.

Across three repository-level evaluations, StateHarness improves task
completion and several measures of change consistency, while adding runtime
overhead relative to the underlying coding agent. It improves full-chain
completion and cleanup consistency on RefactorBench, achieves the highest
mean reward among the compared systems on LHTB at substantially lower
recorded execution cost than LongHorizon-Harness, and increases bug-fix rate
on ChainSWE while reducing residual debt, although pass-to-pass regressions
increase. These results motivate examining both where project-state
maintenance helps and the trade-offs it introduces.

Our work makes three contributions. \textbf{(1)} We introduce an explicit
representation of evolving project state that connects project understanding,
change consequences, and unresolved work throughout a long coding trajectory.
\textbf{(2)} We develop StateHarness, a runtime framework that maintains,
updates, and returns this state to subsequent coding decisions through
lightweight recording, retrieval, and diagnostic mechanisms, without updating
the underlying model. \textbf{(3)} We evaluate StateHarness across refactoring,
long-horizon execution, and sequential bug-fixing settings, characterizing both
its gains in task success and change consistency and its execution-cost
and regression trade-offs.
\section{Related Work}
\label{sec:related-work}

\textbf{Preserving and Reusing Project Understanding. }
Retaining interaction history and preserving reusable knowledge are related
but distinct goals. Context-management methods decide when and what to
compress~\citep{liu2026contexttool,gao2026swemem}, or filter redundant
trajectory content and code
observations~\citep{xiao2025agentdiet,wang2026swepruner,
lindenbauer2025thecomplexitytrap}. External memory and experience-based
methods organize information or distilled strategies for later
reuse~\citep{packer2023memgpt,zhang2025agenticcontextengineeringace,
ouyang2025reasoningbank}. For coding, the retained information may also
need to describe the particular project being developed. ProjectGuard
addresses this through semantic and structural project state that tracks
progressively revealed specifications~\citep{yan2026projectguard}.
StateHarness focuses on the understanding acquired during implementation:
current hypotheses, explanations already ruled out, decision reasons,
and discovered code relationships. It preserves these as editable working
records so that subsequent work can build on earlier analysis, while
fixed-rule updates maintain a compact view without repeatedly summarizing
the full history.

\textbf{Repository Relationships and Change Impact. }
Understanding a repository requires identifying both relevant code and
the consequences of changing it. Task-context tools preserve program
elements relevant to ongoing work, while change-impact analysis connects
source modifications to affected
tests~\citep{kersten2006taskcontext,ren2004chianti}.
Coding agents obtain related context through structure-aware search
and retrieval-augmented
repair~\citep{zhang2024autocoderover,xiong2025thinksearchpatch},
or expose relationships through repository-level code
graphs~\citep{ouyang2025repograph,liu2025codexgraph}.
Keeping that evidence current is another concern: CORVUS synchronizes
file observations as the repository changes~\citep{zheng2026corvus}.
StateHarness addresses the connection between these code relationships
and the agent's ongoing work. After inspecting retrieved locations, the
agent records why a caller, test, or document needs attention, how it has
been handled, and which checks remain. These handling decisions persist
beyond the retrieval step and support review before completion. The focus
is therefore on maintaining task-specific understanding of change
consequences, rather than introducing a new dependency-analysis algorithm.

\textbf{Progress Tracking and Execution Feedback. }
External state can guide what to do next, not just preserve what happened.
Ledger uses tool-derived execution state to inform decisions and mediate
redundant actions, while LongHorizon-Harness uses independently audited
task state to coordinate successive execution
sessions~\citep{wang2026ledger,lh_harness}. Reflective feedback provides
another route from past outcomes to guidance for subsequent
attempts~\citep{shinn2023reflexion}. Execution evidence is also used to
evaluate unsafe skill behavior~\citep{zhuang2026agenttrap}, check action
safety and authorization~\citep{huang2026building,qin2026airguard},
or enforce requirements learned from user
corrections~\citep{zhou2026getting}. Whereas these checks concern safety
or compliance, StateHarness uses feedback to maintain task continuity.
Editing and testing activity prompts updates to the agent's findings and
progress, a working brief returns the current focus and recent results
to subsequent decisions, and occasional diagnosis revisits earlier
recorded evidence. Its central concern is keeping track of where the
work stands and what still needs attention as the project changes.
\section{StateHarness: Maintaining Evolving Project State}
\label{sec:method}

\begin{figure}[t]
\centering
\includegraphics[width=\linewidth]{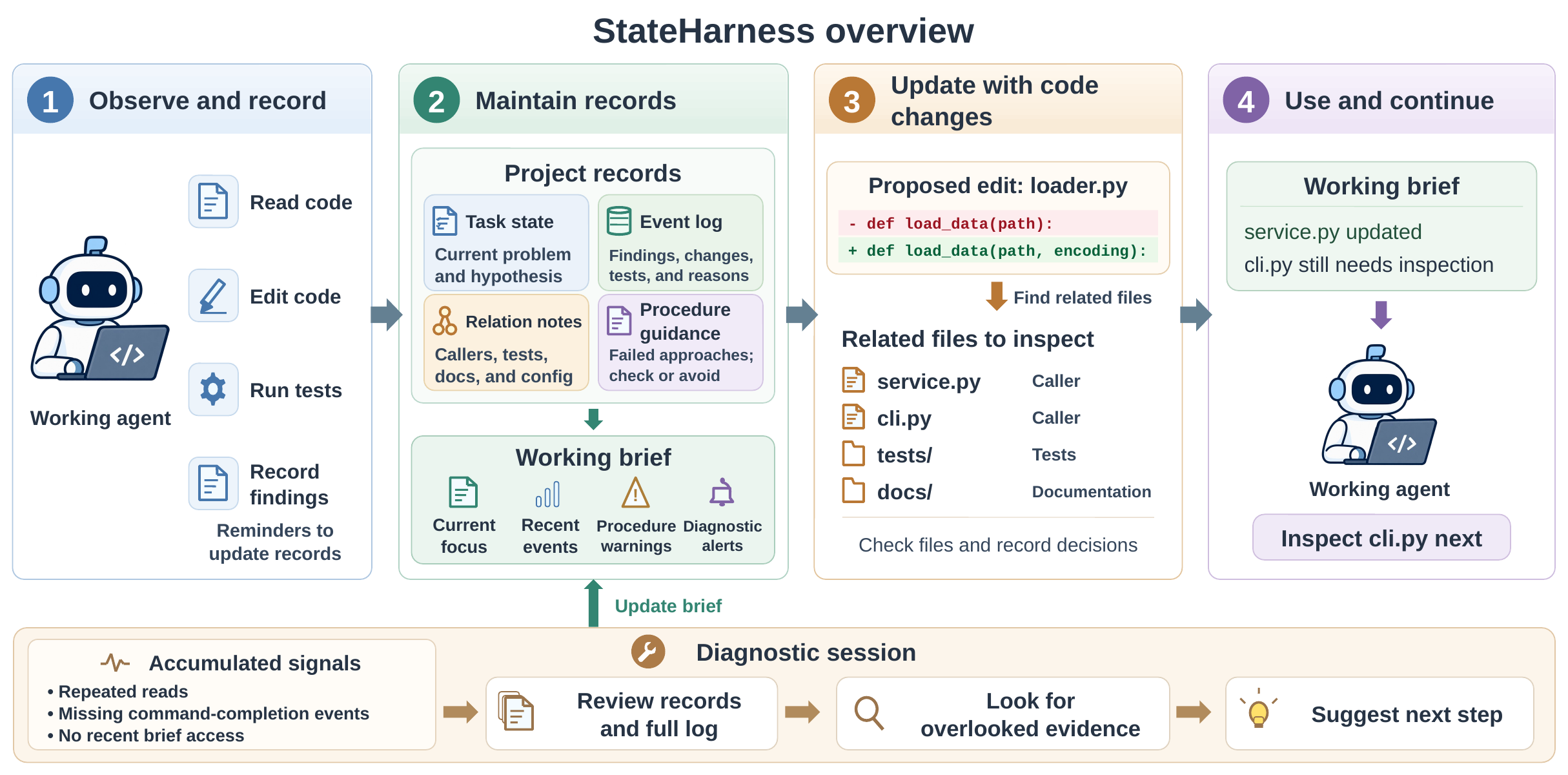}
\caption{StateHarness overview. The working agent maintains persistent project
records, from which fixed rules refresh a compact working brief. Change-impact
retrieval identifies candidate files for inspection when supported edits may
have non-local consequences. Accumulated behavioral signals can trigger a
separate diagnostic session that revisits older recorded evidence and returns
concise suggestions through the brief.}
\label{fig:method-overview}
\end{figure}

StateHarness maintains an evolving, task-centered representation of the
project as coding proceeds. As the working agent reads code, modifies
implementations, and runs tests, it records what it currently understands,
what has changed, which other artifacts may be affected, and what work remains
unfinished. StateHarness then makes this information available to later
decisions through a compact working brief, supplements it with repository
search when changes may have non-local consequences, and revisits older
evidence when execution shows signs that relevant information has been
overlooked. The runtime supports these operations without updating the
underlying model parameters. Figure~\ref{fig:method-overview} summarizes the
interaction between persistent project records, runtime support, and the
working agent.

\subsection{Preserving Project Understanding Acquired During Execution}

The agent develops project understanding incrementally as it works: what the
current problem is, which code is relevant, how components interact, which
explanations remain plausible, and what has already been ruled out.
StateHarness preserves this information outside the conversation so that later
work can build on earlier analysis rather than reconstructing it from the
trajectory.

The current implementation uses four persistent records. \emph{Task state}
stores the initial observation, environment information, and the current
subproblem and hypothesis. An \emph{event log} preserves important findings,
edits, test outcomes, decisions, and reasons for changing direction.
\emph{Relation notes} record discovered connections to callers, tests,
documentation, configuration, or other artifacts that may need to remain
consistent with a change. \emph{Procedure guidance} records approaches that
have repeatedly failed and indicates whether later use should proceed
normally, require explicit verification, or be avoided.

The working agent updates these records as execution proceeds. When its
hypothesis changes, for example, it replaces the current focus in task state
while preserving the earlier reasoning and the reason for the change in the
event log. Observed editing and testing activity can prompt the agent to record
new findings, progress, or relationships. Together, the records preserve the
current task context, accumulated project understanding, and unfinished work
needed by subsequent decisions.

\subsection{Using Existing Understanding in Subsequent Work}

Persistent state is useful only if relevant parts return to the agent when
they matter. A later decision may depend on a cause that was already ruled
out, a caller discovered several steps earlier, or work that was started but
not completed. StateHarness therefore maintains a compact \emph{working brief}
that serves as the routine project-state view for the working agent.

The brief contains the initial environment information, current subproblem and
hypothesis, procedure guidance that affects subsequent actions, recent events,
and diagnostic alerts. The runtime refreshes these sections using fixed rules:
a task-focus update replaces the displayed focus, while additions to the event
log refresh the recent-event window. This synchronization reuses the agent's
recorded content directly rather than invoking another model to summarize the
full history.

Older information remains in the persistent records after it leaves the
brief's recent-event window. The working agent can explicitly revisit these
records when needed, while the diagnostic mechanism described below can also
recover older evidence from them. If the brief has not been accessed for an
extended interval, the runtime returns the current brief to the working agent.
This design keeps routine state compact while retaining a longer history for
later review.

\subsection{Updating Understanding and Progress as Code Changes}

Project state must evolve with the repository. A local edit can affect callers,
tests, documentation, configuration, or other artifacts outside the edited
file, including relationships that the agent has not yet discovered.
StateHarness therefore couples selected code changes with change-impact
retrieval and persistent recording of their consequences.

Before supported operations such as interface changes or source-file
deletions, the runtime searches the repository for locations that may be
affected by the proposed change. The returned locations are candidates rather
than assumed dependencies. The working agent inspects them, determines which
require modification or verification, and records the relevant relationships
and handling decisions.

For example, changing a data-loading function's interface may reveal both a
service caller and a command-line entry point. After updating the service, the
agent can record that the command-line caller still requires inspection; after
the relevant tests run, their outcomes are added to the event log. The working
brief is then refreshed from these records, allowing later decisions to
distinguish completed work from consequences that remain unresolved.

Before termination, the completion protocol asks the agent to revisit the
original request, recorded relationships, relevant tests, and unresolved
warnings. For each relevant artifact, it should either perform the necessary
follow-up or record why no change is required. In this way, discovered
consequences remain connected to their handling status rather than disappearing
after the immediate edit.

\subsection{Recovering Overlooked Information When Execution Falters}

Maintaining project state does not ensure that every relevant record will be
used at the right moment. An agent may repeatedly inspect the same code, lose
track of an earlier finding, or continue an approach that conflicts with
previous evidence. StateHarness therefore monitors behavioral signals such as
repeated reads of the same file, commands that repeatedly lack a normal
completion event, and extended periods without consulting the working brief.

The runtime first responds with ordinary feedback, such as a reminder or the
current brief. If eligible signals continue to accumulate, it starts a fresh
diagnostic session over the task state, procedure guidance, working brief, and
full event log. The diagnostic session looks for evidence of repeated
approaches, stale assumptions, contradictions, or earlier findings that may
have become relevant again, and returns a concise suggested next step when the
recorded evidence supports one.

Diagnostic suggestions are inserted into the working brief and become
available to the working agent on its next decision. For example, an
unresolved caller recorded earlier may eventually leave the recent-event
window; if later execution repeatedly returns to the same part of the code,
diagnosis can recover that earlier evidence from the full log and return it to
the current working view. The working brief therefore supports routine
continuity, while diagnosis provides an occasional mechanism for recovering
older project state when ordinary use of that state appears to break down.
\section{Experiments}
\label{sec:experiments}

We study StateHarness from three complementary perspectives. First, we ask
whether maintaining project state helps a coding agent complete longer
sequences of related work. Second, we examine whether improvements extend
beyond the immediate edit to callers, tests, documentation, and other
artifacts that must remain consistent as the repository changes. Third, we
measure the execution cost of maintaining this state, both relative to the
underlying coding agent and to a more heavily orchestrated long-horizon
harness.

\subsection{Experimental Setup}
\label{sec:experimental-setup}

\textbf{Task settings.}
We evaluate StateHarness in three repository-level settings that emphasize
different aspects of long-horizon coding.

\emph{RefactorBench}~\citep{gautam2025refactorbench} evaluates cross-file
consistency during repository refactoring. We organize the original tasks
into repository-level chains and present the refactorings sequentially while
retaining the benchmark's original requirements and evaluation checks. This
setting measures both completion of the requested transformations and
consistency of related artifacts after a sequence of changes.

\emph{Long-Horizon Terminal-Bench (LHTB)}
~\citep{li2026longhorizonterminalbenchtestinglimitsagents} evaluates sustained
execution over dependent terminal actions. We use five of its six
software-engineering tasks; the remaining task exceeds our execution budget.
We report the benchmark reward for each task and the mean across the evaluated
subset.

\emph{ChainSWE}~\citep{jin2026chainswebenchmarkingcodingagents} evaluates
related software issues on the same evolving repository. We evaluate 37 chains,
presenting the issues in chronological order within a single request. Each
agent starts from the chain's initial repository state, works through the
issues in one continuous execution, and produces one final patch that is
scored issue-by-issue.

\textbf{Models and comparison systems.}
All experiments use Claude Sonnet 5 as the underlying working model. Our
primary baseline is unmodified Claude Code with the same task input and model.
We compare it with StateHarness, which augments the coding agent with
persistent project state, a compact working brief, change-impact retrieval,
runtime reminders, and occasional diagnosis. On LHTB and ChainSWE, we also
compare against LongHorizon-Harness (LH-H)~\citep{lh_harness}, a training-free
framework that coordinates long-horizon execution through manager, executor,
and auditor roles. Holding the working model fixed lets us compare the effect
of the surrounding runtime organization rather than differences in model
capability.

\textbf{Task outcomes and consistency.}
All task outcomes are determined by benchmark tests or external verifiers
rather than agent self-reports. On RefactorBench, we report individual subtask
completion and full-chain completion, together with checks for the requested
refactoring and \emph{cleanup checks} covering dependent imports, call sites,
and documentation. On LHTB, we report the benchmark task reward. On ChainSWE,
we report issue-level bug-fix rate, pass-to-pass (P2P) regressions, and
\emph{residual debt}, a supplementary count of consistency problems that
remain after the chain. The detailed composition of this measure is described
in Appendix~\ref{app:measurement}.

\textbf{Execution cost.}
We report recorded token usage, tool calls, and wall-clock time. These measures
characterize the runtime overhead of StateHarness relative to Claude Code and
its execution cost relative to LongHorizon-Harness. Additional implementation,
scoring, and measurement details are provided in
Appendix~\ref{app:measurement}.

\begin{table}[t]
\centering
\renewcommand{\arraystretch}{1.18}
\setlength{\tabcolsep}{5.5pt}
\caption{\textbf{Overall performance across the three evaluation settings.}
Higher is better for completion, reward, and bug-fix rate; lower is better
for P2P regressions and residual debt. Best results are shown in bold.}
\label{tab:overall-results}
\resizebox{\textwidth}{!}{
\begin{tabular}{lccccccc}
\toprule
\multicolumn{8}{c}{\textbf{RefactorBench}} \\
\cmidrule(lr){1-8}
&
\multicolumn{2}{c}{\textbf{Completion}}
&
\multicolumn{2}{c}{\textbf{Checks}}
&
\multicolumn{3}{c}{\textbf{Cleanup Breakdown}} \\
\cmidrule(lr){2-3}
\cmidrule(lr){4-5}
\cmidrule(lr){6-8}

\textbf{Method}
& \textbf{Subtasks}
& \textbf{Chains}
& \textbf{Refactor}
& \textbf{Cleanup}
& \textbf{Imports}
& \textbf{Calls}
& \textbf{Docs} \\

& {\small $\uparrow$}
& {\small $\uparrow$}
& {\small $\uparrow$}
& {\small $\uparrow$}
& {\small $\uparrow$}
& {\small $\uparrow$}
& {\small $\uparrow$} \\
\midrule

Claude Code
& 77.9
& 29.4
& 94.6
& 87.0
& 82.7
& \textbf{93.3}
& 90.9 \\

\rowcolor{blue!6}
\textbf{StateHarness}
& \textbf{81.4}
& \textbf{35.3}
& \textbf{95.9}
& \textbf{95.4}
& \textbf{94.4}
& \textbf{93.3}
& \textbf{100.0} \\

\midrule[1.2pt]

\multicolumn{8}{c}{\textbf{Long-Horizon Terminal-Bench (LHTB)}} \\
\cmidrule(lr){1-8}

\textbf{Method}
& \textbf{ALP}
& \textbf{GE Audit}
& \textbf{Foldseek}
& \textbf{Unison}
& \textbf{Commit0}
& \multicolumn{2}{c}{\textbf{Mean}} \\
\midrule

Claude Code
& 0.500
& 0.273
& \textbf{0.556}
& 0.167
& 0.639
& \multicolumn{2}{c}{0.427} \\

LongHorizon-Harness
& 0.600
& \textbf{0.364}
& \textbf{0.556}
& 0.167
& \textbf{0.938}
& \multicolumn{2}{c}{0.525} \\

\rowcolor{blue!6}
\textbf{StateHarness}
& \textbf{0.700}
& \textbf{0.364}
& \textbf{0.556}
& \textbf{0.333}
& 0.929
& \multicolumn{2}{c}{\textbf{0.576}} \\

\midrule[1.2pt]

\multicolumn{8}{c}{\textbf{ChainSWE}} \\
\cmidrule(lr){1-8}

\textbf{Method}
& \multicolumn{2}{c}{\textbf{Bug Fix}}
& \multicolumn{2}{c}{\textbf{P2P Regressions}}
& \multicolumn{3}{c}{\textbf{Residual Debt}} \\

& \multicolumn{2}{c}{\small (\%) $\uparrow$}
& \multicolumn{2}{c}{\small $\downarrow$}
& \multicolumn{3}{c}{\small $\downarrow$} \\
\midrule

Claude Code
& \multicolumn{2}{c}{35.1}
& \multicolumn{2}{c}{\textbf{16}}
& \multicolumn{3}{c}{60} \\

LongHorizon-Harness
& \multicolumn{2}{c}{32.5}
& \multicolumn{2}{c}{47}
& \multicolumn{3}{c}{69} \\

\rowcolor{blue!6}
\textbf{StateHarness}
& \multicolumn{2}{c}{\textbf{39.5}}
& \multicolumn{2}{c}{23}
& \multicolumn{3}{c}{\textbf{43}} \\

\bottomrule
\end{tabular}
}
\vspace{2pt}
\end{table}

\subsection{Overall Performance Across Settings}
\label{sec:overall-performance}

Table~\ref{tab:overall-results} summarizes the main results across all three
settings. StateHarness improves the primary task outcome in each setting. On
RefactorBench, full-chain completion increases from 29.4\% to 35.3\%, showing
that the gain extends beyond individual refactoring steps to completion of the
combined sequence. On LHTB, mean reward increases from 0.427 to 0.576;
StateHarness improves over Claude Code on four of the five evaluated tasks and
ties on Foldseek. On ChainSWE, issue-level bug-fix rate increases from 35.1\%
to 39.5\%.

The consistency results are less uniform. On RefactorBench, the requested
refactoring checks improve only modestly, whereas cleanup checks show a larger
gain. On ChainSWE, reported residual debt decreases from 60 to 43, but
pass-to-pass regressions increase from 16 to 23. The latter result is important:
maintaining additional project state does not uniformly preserve previously
passing behavior, even when more requested issues are resolved. We therefore
treat task completion and change consistency as related but distinct outcomes.

StateHarness also compares favorably with LongHorizon-Harness in the two
settings where both are evaluated. On LHTB, its mean reward is 0.576 compared
with 0.525 for LongHorizon-Harness. The comparison varies by task:
StateHarness performs better on ALP and Unison, ties on GE Audit and Foldseek,
and is slightly lower on Commit0. On ChainSWE, StateHarness achieves a higher
bug-fix rate (39.5\% versus 32.5\%), with fewer P2P regressions and lower
reported residual debt. These results suggest that the benefit is not tied to
a single task structure: it appears across sequential refactoring, long
terminal workflows, and multi-issue repository maintenance.

\subsection{Consistency Beyond the Immediate Change}
\label{sec:change-consistency}

\begin{figure}[t]
\centering
\includegraphics[width=\linewidth]{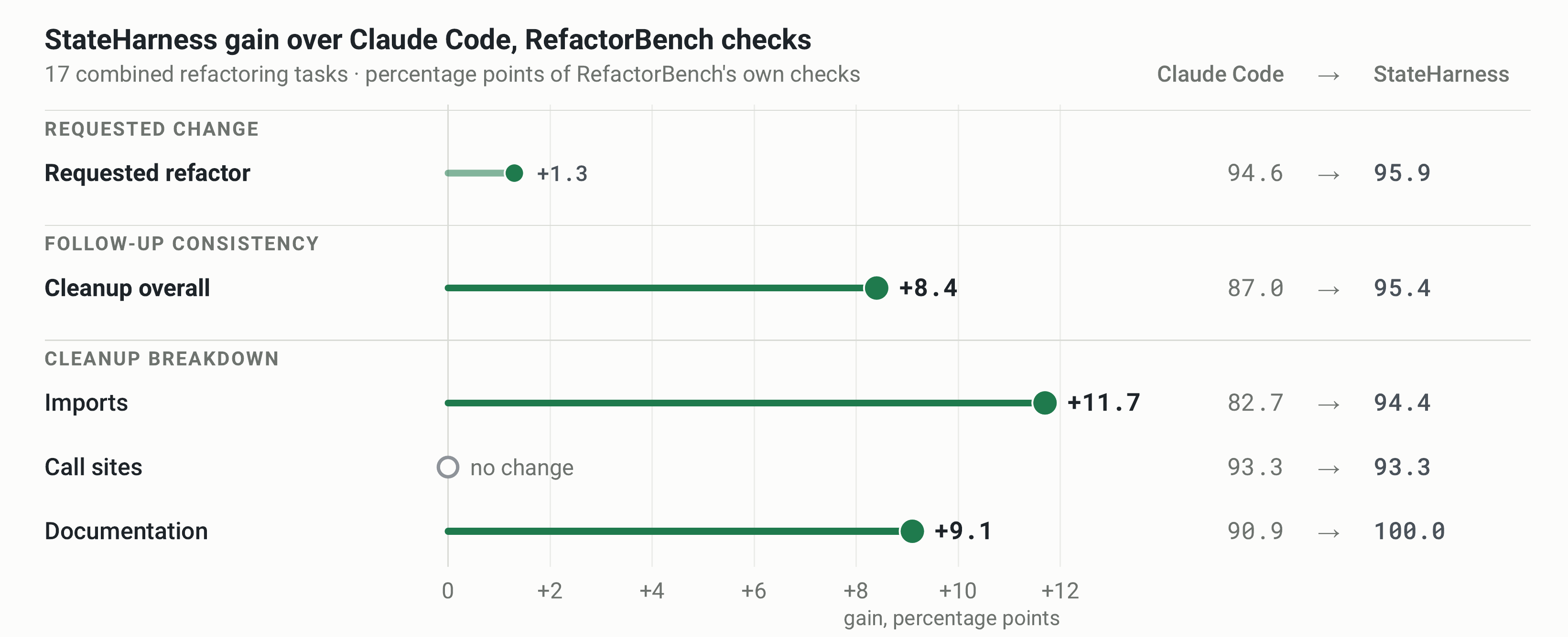}
\caption{\textbf{RefactorBench consistency breakdown.}
StateHarness produces a modest improvement on checks for the requested
refactoring itself, while larger gains appear in cleanup checks for related
artifacts. The largest improvements occur for import and documentation updates;
call-site performance is unchanged.}
\label{fig:refactor-consistency}
\end{figure}

RefactorBench lets us separate the requested transformation from the
follow-up work needed to keep related artifacts consistent.
Figure~\ref{fig:refactor-consistency} shows the corresponding gains over
Claude Code.

The larger difference appears after the immediate edit: cleanup accuracy
improves by 8.4 percentage points, compared with 1.3 points for the requested
refactoring checks. The cleanup gain is driven by imports (+11.7 points) and
documentation (+9.1 points), while call-site performance is unchanged.
StateHarness therefore improves several forms of follow-up consistency rather
than uniformly improving every related artifact.

The distinction matters because the requested edit and its downstream
consequences place different demands on the agent. Completing the immediate
refactoring primarily requires modifying the target implementation correctly;
cleanup additionally requires remembering which surrounding artifacts were
affected and returning to them after the local edit has been made. The larger
gains on imports and documentation are therefore consistent with the intended
role of StateHarness in keeping change consequences connected to later work.
At the same time, the unchanged call-site result shows that this benefit is
not universal across all related artifacts.

\begin{table}[t]
\centering
\renewcommand{\arraystretch}{1.18}
\setlength{\tabcolsep}{6.5pt}
\caption{\textbf{Execution cost across evaluation settings.}
Values are averages under the corresponding setting. Lower is better for
all cost measures.}
\label{tab:execution-cost}
\begin{tabular}{llccc}
\toprule
\textbf{Setting} & \textbf{Method}
& \textbf{Recorded tokens (K)}
& \textbf{Tool calls}
& \textbf{Time (s)} \\
\midrule

\multirow{2}{*}{RefactorBench}
& Claude Code & 25.6 & 81 & 383 \\
& \textbf{StateHarness} & 33.1 & 110 & 529 \\
\addlinespace[2pt] \midrule

\multirow{3}{*}{ChainSWE}
& Claude Code & 24.9 & 75 & 300 \\
& \textbf{StateHarness} & 35.3 & 96 & 383 \\
& LongHorizon-Harness & 208.0 & 272 & 2214 \\

\bottomrule
\end{tabular}
\end{table}
\subsection{Execution Cost and Performance Trade-offs}
\label{sec:execution-cost}

StateHarness incurs additional runtime work relative to the underlying Claude
Code agent. On RefactorBench, recorded token usage increases from 25.6K to
33.1K, tool calls from 81 to 110, and wall-clock time from 383 to 529 seconds.
ChainSWE shows the same direction: StateHarness uses 35.3K recorded tokens,
96 tool calls, and 383 seconds, compared with 24.9K, 75, and 300 seconds for
Claude Code. Project-state maintenance therefore introduces measurable
runtime overhead rather than reducing the execution cost of the base agent.

The comparison with LongHorizon-Harness shows a different trade-off. On
ChainSWE, LongHorizon-Harness uses 208.0K recorded tokens, 272 tool calls, and
2214 seconds, while StateHarness uses 35.3K tokens, 96 calls, and 383 seconds
and achieves the higher bug-fix rate. The same qualitative pattern appears on
LHTB: LongHorizon-Harness uses more wall-clock time and tool calls on every
evaluated task, while StateHarness achieves the higher mean reward. Commit0 is
the main task-level exception in performance, where LongHorizon-Harness
obtains a slightly higher reward (0.938 versus 0.929) at substantially greater
runtime cost.

Because auxiliary harness sessions are not fully represented in the unified
token accounting, all token comparisons are reported as \emph{recorded token
usage}; Appendix~\ref{app:measurement} describes the accounting in detail.
Taken together, these results place StateHarness between the two comparison
points: it adds runtime work relative to the base coding agent, but is
substantially lighter in the recorded measurements than the multi-role
LongHorizon-Harness while achieving stronger overall task performance.

\subsection{Case Studies}
\label{sec:case-studies}

Aggregate results do not show how information preserved earlier in a task
affects later coding decisions. We therefore examine representative examples
from ChainSWE and LHTB. They illustrate three outcomes: earlier information
can motivate additional checking, an unmet requirement can lead to follow-up
work, and a known problem can still remain unresolved when the agent stops.

\begin{figure}[t]
\centering
\includegraphics[width=\linewidth]{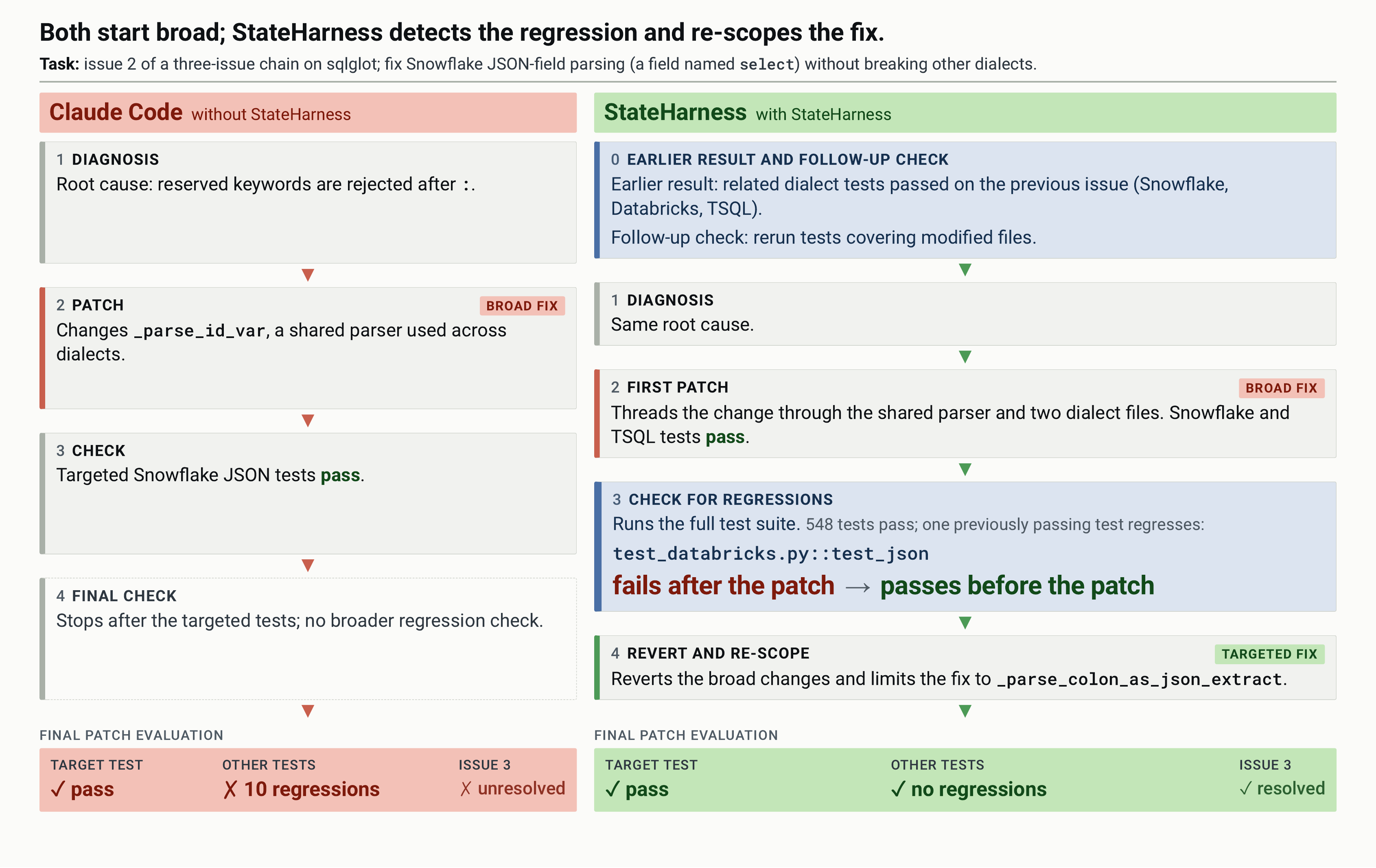}
\caption{\textbf{Regression checking changes the repair trajectory.}
Both agents identify the same root cause and initially make broad changes to
shared parsing behavior. Claude Code stops after the targeted tests pass.
StateHarness performs additional regression checking, detects a failure in a
previously passing Databricks test, and narrows the repair to the JSON-path
parser.}
\label{fig:sqlglot-case}
\end{figure}

\textbf{Regression checking changes the repair trajectory.}
Figure~\ref{fig:sqlglot-case} shows issue 2 of a three-issue ChainSWE example
in \texttt{sqlglot}. Claude Code and StateHarness identify the same root
cause---reserved keywords are rejected after the JSON-field separator---and
both initially modify shared parsing behavior. Claude Code runs the targeted
Snowflake tests, observes that they pass, and stops. Later evaluation reveals
ten regressions in other tests.

StateHarness also begins with a broad repair, but it retains the earlier result
that related dialect tests had passed and is instructed to rerun tests covering
modified files. The additional check reveals that
\texttt{test\_databricks.py::test\_json}, which passed before the patch, now
fails. The agent reverts the broad changes and restricts the repair to the
JSON-path parser. The final patch passes the target test without introducing
regressions, and the agent continues to resolve issue 3. This example shows
how information from earlier work can remain useful after the immediate target
test has already passed.

\textbf{Turning an unmet requirement into follow-up work.}
In \texttt{commit0-multilib-tdd}, the task requires complete implementations,
passing self-written tests, and at least 90\% branch coverage. After the
initial implementation, 435 tests pass, but coverage reaches only 85.9\%.
The agent records the remaining coverage gap and the largest uncovered
regions, then works on those branches. Coverage subsequently reaches 98.6\%,
and the agent checks the requirement again before finishing. Here, a known
shortfall is carried forward, acted on, and verified.

\textbf{Remembering a problem does not guarantee fixing it.}
The opposite outcome appears in \texttt{alp-paper-reproduction}. The agent
produces a working codec but observes compression values of 56.67 and
56.63 bits/value against targets of 57.50 and 56.75. It records the mismatch
and notes that further calibration may be needed, but ends without another
iteration. The problem is therefore remembered but not resolved.

Together, these examples separate three steps: remembering an unresolved
problem, acting on it, and verifying the result. StateHarness can help carry
information across all three steps, but recording a problem does not force the
agent to resolve it before stopping. This distinction mirrors the role of StateHarness more generally: it keeps
relevant information available to later decisions, but the working agent
still determines how and when that information is acted upon.

\section{Conclusion}
\label{sec:conclusion}
We introduced StateHarness, a training-free runtime framework that maintains evolving project state during long-horizon coding: task understanding, decisions, discovered code relationships, and unresolved work, so later execution builds on earlier analysis as the repository changes. Across three repository-level settings it improves task completion and several measures of change consistency. On RefactorBench the largest gains are in chained completion and cleanup of related artifacts; on LHTB it attains higher mean reward than Claude Code and LongHorizon-Harness at substantially lower recorded cost than the latter; on ChainSWE it resolves more issues with lower residual debt but more pass-to-pass regressions. Explicit project-state maintenance thus supports more coherent long-horizon coding, at the cost of runtime overhead and trade-offs across consistency measures. The case studies also expose a limitation: remembering an unresolved problem enables later checking but does not guarantee the agent resolves it before stopping.

\section*{AI Use Statement}

In this work, we used generative AI tools for language polishing and editing
assistance. Generative AI models are also an explicit component of our
experimental setting, where coding agents powered by large language models are
evaluated under different runtime configurations; these uses are described in
the methodology and experimental setup. We did not use generative AI to
generate ground-truth labels or modify benchmark evaluation criteria. All
AI-assisted text and reported experimental outputs were reviewed by the
authors. We take full responsibility for the final content of this work,
including its claims, analyses, and artifacts.

\section*{Ethics Statement}

This work studies runtime mechanisms for supporting long-horizon coding agents.
Our experiments use public software-engineering benchmarks and their associated
code repositories, and agent execution is restricted to benchmark or isolated
development environments. The study does not involve human subjects, private
user data, or personally identifiable information. StateHarness does not
modify or retrain the underlying model. It maintains external project records,
provides runtime feedback during coding, and can invoke a separate diagnostic
session when behavioral signals accumulate. We provide the implementation and
evaluation artifacts to support reproducibility and further research on
long-horizon coding-agent systems.

\section*{Reproducibility Statement}

To facilitate reproducibility, Section~\ref{sec:method} describes the
StateHarness design and project-state representation, while
Section~\ref{sec:experiments} specifies the evaluation settings, comparison
systems, and outcome measures. Appendices~\ref{app:runtime},
\ref{app:measurement}, and~\ref{app:prompts} document the persistent records,
runtime update and retrieval rules, diagnostic conditions, prompt delivery,
implementation versions, benchmark-specific scoring, and execution-cost
accounting. StateHarness is training-free and does not modify the parameters
of the underlying coding model. The exact protocol and template snapshots,
their hashes, example records, and evaluation artifacts are included in the
accompanying reproducibility materials. Reported task outcomes are determined
by benchmark tests or external evaluators rather than by agent self-reports.

\bibliography{references}
\bibliographystyle{preprint}

\clearpage
\appendix
\section{StateHarness Details}
\label{app:runtime}

This appendix provides implementation details for the four functions described
in Section~\ref{sec:method}: how project understanding is recorded, how
relevant state is returned to current work, how that state is updated as the
repository changes, and how older evidence is reconsidered when ordinary state
use breaks down. We conclude with a constructed example that follows one
change through these functions.

\subsection{Persistent Project Records}
\label{app:records}

StateHarness represents task-relevant project understanding through a small
set of persistent records. These records are authored and updated by the
working agent as execution proceeds. Table~\ref{tab:record-mapping}
summarizes their roles.

The initial task state records the available task observation and environment
information. For failure-driven tasks, this includes the reproduction command,
its outcome, and the observed failure category. The \emph{Current Focus} is a
single line of at most 200 characters that combines the current subproblem
with the agent's working hypothesis. When the hypothesis changes, the previous
reasoning remains available in the event log. The task goal originates from
the user request; although the task-state template does not require a separate
goal field, the agent may preserve goal-related information as a decision
event.

\begin{table}[htbp]
\centering\small
\caption{\textbf{Persistent representations of task-relevant project state.}
The working agent authors the first four records; the working brief is a
selective view returned to subsequent decisions.}
\label{tab:record-mapping}
\begin{tabularx}{\linewidth}{
@{}
>{\raggedright\arraybackslash}p{.19\linewidth}
>{\raggedright\arraybackslash}X
>{\raggedright\arraybackslash}p{.29\linewidth}
@{}}
\toprule
\textbf{Record}
& \textbf{What changes during the task}
& \textbf{Stored representation} \\
\midrule

Task state
& Record the initial observation and environment; replace the current focus
when the active subproblem or hypothesis changes.
& \path{task_state.md} \\

Event log
& Append important changes, tests, findings, decisions, and rejected
assumptions together with their reasons.
& \path{breakthrough_log.md} \\

Relation notes
& Record discovered relationships to callers, tests, documentation,
configuration, or schemas.
& \path{file_relation_graph.md} \\

Procedure guidance
& Maintain a named procedure's invocation history, latest observed failure,
usage policy, and rationale.
& \path{skill_states.json} \\

Working brief
& Refresh the current focus, relevant procedure guidance, recent events, and
diagnostic alerts returned to the working agent.
& \path{context_brief.md} \\

\bottomrule
\end{tabularx}
\end{table}

Procedure guidance is created for procedures named by the agent rather than
from a fixed registry of every tool call. The current lifecycle assigns
\texttt{use} after a first observed failure, \texttt{verify} after a second
failure with the same cause, and \texttt{avoid} after a third failure or a
confirmed repeated pattern. These policies respectively indicate normal use,
explicit inspection of the output before proceeding, and selection of an
alternative unless an exception is justified. A state-corrupting or
unrecoverable failure can instead justify \texttt{freeze}, which prohibits
further use of the named procedure within the current task. These policies
guide procedure selection but do not modify tool permissions.

The working agent assigns the failure category used to update procedure
guidance. Initialization and instruction delivery are described in
Appendices~\ref{app:initialization} and~\ref{app:protocol-delivery}. The exact
templates and event vocabulary are included in the supplied
\path{Reproducibility/} materials.

\subsection{Maintaining the Working Brief}
\label{app:anchoring}

The persistent records are useful only if relevant information returns to the
agent's current work. StateHarness therefore maintains a compact working brief
that exposes the current focus, relevant procedure guidance, recent events,
and diagnostic alerts without repeatedly summarizing the full interaction
history.

As execution proceeds, the runtime uses observed editing and testing activity
to remind the working agent to keep the records current. After three observed
edits to non-state files, it requests updates to modification events, the
current focus when needed, and newly discovered relations; at most two such
edit reminders are issued for the stored count. Every three recognized test
invocations similarly request an outcome record. These reminders prompt
record maintenance rather than interpreting the result of an edit or test.

For recognized record writes, synchronization updates the corresponding
section of the working brief. A focus change replaces the displayed focus, an
update to procedure guidance refreshes the guidance shown in the brief, and
an event-log write refreshes the recent-event window. The initialized
environment remains available in task state, while later environmental
changes can be recorded as events.

The implementation monitors supported edit operations and recognized shell
writes. Unsupported filesystem mutations are not continuously tracked.
Feedback generated from a recognized operation is returned with the next
model request.

Several lightweight checks preserve the structure and continuity of the
records. Before registered edit tools overwrite the event log, history
protection can reject removal of previously recognized entries. After
supported writes, validation checks the event tags, current-focus line, and
procedure-policy fields and counters. Formatting errors request a later
correction without undoing the write. These checks preserve record structure and history; they do not assess
whether the agent's interpretation of the recorded evidence is correct.

\subsection{Discovering and Revisiting Change Impact}
\label{app:impact}

StateHarness supplements the agent's accumulated understanding with
repository search when a supported operation may have consequences outside
the edited location. Retrieval begins from the proposed operation and the
current repository contents. Where available, syntax-tree information is used
to identify affected functions or classes, with textual matching as a
fallback.

The current implementation triggers this search for supported signature
changes, whole-file replacements that expose changed or removed definitions,
and source-file deletions. Body-only edits normally do not trigger this
retrieval path. Search returns bounded groups of candidate source, test, and
documentation locations, with current coverage for Python and Cython source
and supported plain-text documentation.

Retrieved locations are candidates rather than automatically established
dependencies. The working agent determines whether a returned source file,
test, document, configuration entry, or schema is actually related to the
change and whether follow-up work is necessary. Relation notes preserve these
judgments for later use. Retrieval itself operates from the proposed change
and repository contents; it does not use the existing relation notes as its
search index.

Before completion, the agent is asked to revisit the recorded relationships,
rerun the original reproduction when feasible, and execute relevant tests.
For each relevant artifact, the agent should either make the corresponding
change or record why no change is required. The completion protocol also
advises reserving a full test-suite run for the final stage and running it at
most once. This is guidance to the working agent rather than a runtime
restriction; additional test runs are not blocked. The handling status of
discovered consequences remains in the project records for review before
completion.

\subsection{Recovering Overlooked Information}
\label{app:recovery}

The working brief normally provides the information needed for continued
execution, but relevant evidence can still be overlooked. StateHarness
therefore monitors behavioral signals that indicate when older project state
may need to be revisited. These signals are behavioral proxies derived from
tool use. The current implementation monitors repeated reads, missing
command-completion events, and extended intervals without brief access, while
excluding StateHarness files and recognized exploration operations from the
corresponding counters.

Signals contribute to a shared diagnostic score. An eligible signal attempts
diagnosis once the score reaches two, provided that no diagnostic cooldown is
active and the required diagnostic inputs are available. Otherwise the
runtime returns ordinary feedback such as a reminder or the current brief.
Cooldown decreases with monitored tool calls; reaching zero does not itself
invoke diagnosis. Table~\ref{tab:signal-transitions} summarizes the default
transition rules.

\begin{table}[htbp]
\centering\small
\caption{\textbf{Default feedback and diagnostic transitions.}
Unmentioned counters retain their current values. Test completions and
recognized progress tags act as progress proxies.}
\label{tab:signal-transitions}
\begin{tabularx}{\linewidth}{
@{}
>{\raggedright\arraybackslash}p{.27\linewidth}
>{\raggedright\arraybackslash}X
@{}}
\toprule
\textbf{Observed condition}
& \textbf{Response and state change} \\
\midrule

Missing command-completion events
& At the next shell call, count a missing normal completion callback.
Three such events produce a signal and reset the streak; a normal shell
completion clears the streak. \\

Repeated reads of one path
& Three counted reads produce a signal and reset that path's count.
A proposed edit to the path clears the count before the edit outcome is known. \\

Working brief neglected for 25 calls
& Deliver the brief and restart the window. Add diagnostic score only when no
non-state edit or recognized test was proposed during that window.
Recognized brief access restarts the window. \\

Recognized test completion
& Clear diagnostic score and diagnostic alerts on the normal post-event, without
parsing whether the test passed. A proposed test alone does not perform
this reset. \\

Recognized progress tag
& Synchronization clears the diagnostic score and alerts when the latest event
indicates recognized progress, such as a passed test or an established root
cause. \\

Successful nonblank diagnosis
& Replace diagnostic alerts, clear diagnostic score and the brief-neglect count,
start a 15-call cooldown, and return the updated brief. \\

Diagnostic failure
& Retain diagnostic score and prior alerts, start a five-call cooldown, and return
ordinary feedback. \\

\bottomrule
\end{tabularx}
\end{table}

A diagnostic session receives the triggering signal, task state, procedure
guidance, full event log, and current working brief. Relation notes and the
complete tool transcript are not supplied as separate inputs; task and goal
information reaches diagnosis through the maintained task state.

The diagnostic session uses a separately configured model and is instructed
to cite recorded evidence and suggest a next step without modifying the
repository. It returns at most three alerts. Successful alerts replace the
diagnostic-alert section of the working brief; if diagnosis fails, the runtime
retains the previous alerts and returns ordinary feedback.
Appendix~\ref{app:prompts} provides the exact prompt and delivery context.

\subsection{Worked Example}
\label{app:worked}

Table~\ref{tab:method-example} illustrates how the four functions interact in
a constructed example. The task changes
\texttt{load\_config(path)} so that it requires an encoding argument. An
initial loader test rejects the extra argument; both a service and a CLI call
the function, and a README contains an example of its use. The example is
intended to clarify information flow rather than serve as an experimental
trajectory. A serialized intermediate snapshot is included in
\path{Reproducibility/examples/}.

\begin{table}[H]
\centering\small
\caption{\textbf{One change through the four StateHarness functions.}
The example shows how project understanding is recorded, returned to current
work, updated after code changes, and later recovered when relevant evidence
falls out of the recent working view.}
\label{tab:method-example}
\renewcommand{\arraystretch}{1.13}
\begin{tabularx}{\linewidth}{
@{}
>{\raggedright\arraybackslash}p{.19\linewidth}
>{\raggedright\arraybackslash}X
>{\raggedright\arraybackslash}X
@{}}
\toprule
\textbf{Function}
& \textbf{Recorded project state}
& \textbf{Subsequent use} \\
\midrule

Preserve understanding
& Retrieval identifies the service, CLI, test, and README as potentially
affected by the loader signature change. The event log records that the
loader and service have been updated, while relation notes preserve the
remaining caller and documentation relationships.
& The discovered scope remains available after the immediate edit, including
the still-unchecked CLI caller. \\

Return current state
& The current focus becomes ``Checking \texttt{loader.py}; I suspect encoding
is still handled incorrectly.'' Recent modification and decision events are
copied into the working brief together with relevant procedure guidance.
& Subsequent model requests receive the active hypothesis and recent progress
without rereading the full project history. \\

Update after changes
& Further edits push the earlier CLI decision outside the recent-event window,
but the decision remains in the full event log. Later state records show that
the loader and service have been updated while the CLI still requires review.
& The compact working view follows recent work, while older project evidence
remains recoverable from persistent state. \\

Recover overlooked evidence
& Repeated reads of \texttt{loader.py} raise diagnostic score to the diagnostic
threshold. Diagnosis revisits the full event log and surfaces the deferred
CLI inspection as overlooked evidence.
& The alert returns the CLI to the current working view. After the CLI is
updated and the relevant regression test passes, the new result is recorded
and ordinary execution continues. \\

\bottomrule
\end{tabularx}
\end{table}
\section{Experimental and Measurement Details}
\label{app:measurement}

This appendix provides additional details on the comparison systems, task
boundaries, implementation versions, outcome measurement, and execution-cost
accounting used in the experiments.

\subsection{Comparison Systems}
\label{app:configurations}

Our primary comparison is between the ordinary Claude Code agent and the full
StateHarness configuration. The two configurations use the same working model
and receive the same task input within each evaluation. StateHarness augments
the underlying agent with persistent project records, working-brief
maintenance, repository retrieval, runtime reminders, lightweight record
validation, and on-demand diagnosis. Table~\ref{tab:configuration-matrix}
summarizes the difference between the two configurations.

\begin{table}[htbp]
\centering\small
\caption{\textbf{Runtime components in the ordinary-agent and StateHarness
configurations.} The working model and task input are matched within each
comparison.}
\label{tab:configuration-matrix}
\begin{tabularx}{\linewidth}{@{}Xcc@{}}
\toprule
\textbf{Component}
& \textbf{Claude Code}
& \textbf{StateHarness} \\
\midrule
Persistent project records
& No & Yes \\
Working-brief synchronization
& No & Yes \\
Recording reminders and behavioral signals
& No & Yes \\
Repository change-impact retrieval
& No & Yes \\
On-demand diagnostic session
& No & Yes \\
Record-structure and history checks
& No & Yes \\
\bottomrule
\end{tabularx}
\end{table}

On LHTB and ChainSWE, we additionally compare against
LongHorizon-Harness. LongHorizon-Harness decomposes execution across a
manager, fresh executor sessions, and an auditor that evaluates progress
before subsequent work proceeds. It is therefore a distinct multi-role
long-horizon solver rather than an analogue of the occasional diagnostic
session used by StateHarness.

\subsection{Initialization and Task Boundaries}
\label{app:initialization}

For each independent StateHarness run, the runtime begins from fresh project
records and runtime counters. The initial request provides the task
description together with the repository and execution information available
for that benchmark. The working agent is asked to establish an initial
observation and current focus before maintaining the records during subsequent
work. For failure-driven tasks, this observation can include a reproduction
command and its observed result. Initialization does not prevent the agent
from beginning repository work if such a reproduction is unavailable.

Procedure guidance is stored separately from the other project records. The
supplied template describes its schema and lifecycle, while the runtime starts
from an empty task-local store when that initialization step is performed.
Because procedure entries are created from failures observed during execution,
the store need not contain a predefined registry of tools or procedures.

\paragraph{ChainSWE execution boundary.}
In ChainSWE, all issues belonging to one chain are presented in chronological
order within a single request. The agent begins from the chain's initial
repository state, works in one continuous session, and produces one final
patch that is evaluated issue-by-issue. StateHarness records therefore evolve
within the same execution as successive issues are addressed. The experiment
does not evaluate transfer of project state between separate requests or
independent agent sessions.

\paragraph{Retries and partial attempts.}
A timed-out attempt can still export an available partial patch for benchmark
scoring while retaining its timeout status. Setup or export failures are
tracked separately from benchmark judgments on successfully exported patches.
When an attempt is retried, failed-attempt artifacts are archived and the
preceding StateHarness record snapshot is restored when available rather than
unconditionally continuing from the failed attempt's partial state.

\subsection{Models and Implementation Versions}
\label{app:version}

All reported experiments use Claude Sonnet 5 as the working model. The
StateHarness implementation used for the protocol and project-state templates
is pinned to commit \texttt{72611b1}. The ChainSWE combined-task construction
and local scorer are specified from commit \texttt{ba0100f}, and the
integration uses Claude Code version \texttt{2.1.261}.

The diagnostic-session configuration is recorded separately from the working
agent configuration. The released run metadata also records the task baseline,
StateHarness version, and scorer version associated with each result.

\subsection{Task Outcomes}

All reported task outcomes are determined from benchmark tests or external
verifiers rather than from the agent's final prose or its internal project
records.

For RefactorBench, we retain the benchmark's original checks while presenting
related refactoring tasks sequentially at the repository level. We report
subtask completion, full-chain completion, requested-refactoring checks, and
cleanup checks covering dependent imports, call sites, and documentation.

For LHTB, task outcomes are produced by the benchmark verifier through Harbor.
We report the reward assigned by the corresponding task verifier and the mean
over the five software-engineering tasks included in our evaluation.

For ChainSWE, the agent produces one patch for the full issue chain. The local
scorer applies that patch to the chain baseline in a fresh scoring environment
and evaluates the issues in chronological order. Benchmark test patches are
applied cumulatively, and each issue is evaluated with its associated test
command. An issue is counted as resolved when the relevant observed
fail-to-pass tests pass and no observed pass-to-pass test fails or errors.
Tests expected by the benchmark but absent from the parsed log are tracked as
missing rather than silently counted as passing. We report the fraction of
issues resolved, pass-to-pass regressions, and residual debt remaining after
the chain.

The residual-debt measure includes unresolved repository inconsistencies such
as stale callers after interface changes, remaining regressions, outdated
documentation, and, for systems that maintain explicit project records, stale
recorded state. Because the last category applies only to methods that maintain
such records, residual debt should be interpreted as a supplementary
consistency measure rather than as a method-independent task-success metric.

\subsection{Execution-Cost Accounting}

We report recorded token usage, tool calls, and wall-clock time to characterize
execution cost. Token counts use the model-usage information available from
the corresponding runtime. Diagnostic sessions are not fully included in the
unified token accounting, so throughout the paper we refer to these values as
\emph{recorded token usage} rather than total inference cost.

Where runtime-injected text is measured separately, its length is estimated
using four characters per token and is not added as an exact charge to the
model-usage totals. Tool-call and wall-clock measurements include the observed
working-agent actions, including additional test runs beyond protocol
guidance. In the \texttt{sqlglot} case study, both full-suite runs remain in
the trajectory and are included in the reported execution accounting.

Because the compared harnesses organize model calls and auxiliary sessions
differently, these measurements characterize observed execution cost rather
than a compute-matched comparison.
\section{Prompts and Input Delivery}
\label{app:prompts}

This appendix describes how StateHarness instructions are delivered to the
working agent and diagnostic session. The paper summarizes the behaviorally
relevant instructions here; the exact prompt and protocol snapshots used in
the experiments are included in the accompanying reproducibility materials.

\subsection{Input Delivery}
\label{app:protocol-delivery}

StateHarness uses three forms of instruction delivery: text included directly
in a model request, explicit requests to read protocol files, and supporting
files made available in the StateHarness directory. We distinguish these
forms because they expose instructions to the working agent in different ways.

\begin{table}[htbp]
\centering\small
\caption{\textbf{Instruction delivery in the StateHarness integration.}
Availability indicates that a file can be accessed by the agent; explicit
delivery indicates that its contents or a request to read it enter the
execution directly.}
\label{tab:input-delivery}
\begin{tabularx}{\linewidth}{
@{}
>{\raggedright\arraybackslash}p{.25\linewidth}
>{\raggedright\arraybackslash}p{.22\linewidth}
>{\raggedright\arraybackslash}X
@{}}
\toprule
\textbf{Material}
& \textbf{Delivery}
& \textbf{Role} \\
\midrule

Working-agent startup instructions
& Included in the initial request
& Provide environment information, identify the project-state directory,
request initialization and completion checks, and specify when the working
agent should update project records. \\

Initialization and completion protocols
& Explicitly requested reads
& Establish the initial task observation and current focus, and later request
review of recorded relationships, unresolved warnings, and relevant tests
before termination. \\

Procedure-guidance lifecycle
& Available as protocol files
& Specifies how repeated procedure failures can produce
\texttt{use}, \texttt{verify}, \texttt{avoid}, or \texttt{freeze} guidance.
These files are not independently injected on every procedure failure. \\

Record templates and maintenance protocols
& Available as supporting files
& Define the persistent record formats and additional maintenance behavior.
The working records are initialized from these templates rather than requiring
the agent to reread the templates during ordinary execution. \\

Diagnostic instructions
& Supplied directly to a fresh diagnostic session
& Request evidence-backed identification of overlooked information or
unproductive execution patterns using the triggering signal, task state,
procedure guidance, full event log, and current working brief. \\

\bottomrule
\end{tabularx}
\end{table}

The exact protocol and template snapshots used for the reported experiments
are preserved under \path{Reproducibility/protocols/} and
\path{Reproducibility/templates/}. Their source locations and SHA-256 hashes
are recorded in \path{Reproducibility/manifest.json}. Filled examples under
\path{Reproducibility/examples/} instantiate the worked example in
Appendix~\ref{app:worked}.

\subsection{Working-Agent Instructions}
\label{app:working-prompt}

The startup instructions establish how the working agent interacts with the
persistent project state. At launch, the runtime supplies the locations of the
repository, project-state files, and protocols together with known environment
information. The agent is asked to preserve these environment facts rather
than rediscover them through unnecessary probing.

The working agent is then instructed to initialize the task state and maintain
the following records during execution:

\begin{itemize}
    \item append important findings, code changes, test outcomes, decisions,
    rejected assumptions, and dependency changes to the event log;
    \item replace the current focus when the active subproblem or hypothesis
    changes;
    \item record newly discovered cross-file relationships, such as callers,
    tests, and documentation that may need to remain consistent with a change;
    \item use the working brief as the compact project-state view returned to
    subsequent decisions; and
    \item run focused tests for modified behavior and record their summarized
    outcomes, reserving a full test-suite invocation for the final stage when
    appropriate.
\end{itemize}

Before termination, the completion protocol asks the working agent to revisit
the original task, recorded relationships, unresolved warnings, and relevant
verification results. The instructions also advise reserving a full
test-suite run for the final stage and running it at most once. These are
guidance to the working agent rather than runtime restrictions: additional
test runs are not blocked, and an unresolved warning does not mechanically
prevent completion.

\subsection{Diagnostic-Session Instructions}
\label{app:diagnostic-prompt}

When the behavioral signals described in Appendix~\ref{app:recovery} reach
the diagnostic threshold, StateHarness starts a fresh diagnostic session. The
session receives the triggering signal, task state, procedure guidance, full
event log, and current working brief. Task and goal information reaches the
session through the maintained task state.

The diagnostic session is asked to inspect older recorded evidence for
repeated approaches, stale assumptions, contradictions, or findings that may
have become relevant again. It may return at most three evidence-backed
alerts, each with a concise suggested next step. If the available evidence
does not support a diagnosis, it returns no alert. Successful alerts are
inserted into the working brief and become available to the working agent on
its next decision. The exact diagnostic prompt used in the experiments is
included in the reproducibility materials.

\end{document}